\documentstyle{article}
\makeatletter
\renewcommand{\theequation}{\thesection.\arabic{equation}}
\@addtoreset{equation}{section}
\def\eqnarray{%
\stepcounter{equation}%
\let\@currentlabel=\theequation
\global\@eqnswtrue
\global\@eqcnt\z@
\tabskip\@centering
\let\\=\@eqncr
$$\halign to \displaywidth\bgroup\@eqnsel\hskip\@centering
$\displaystyle\tabskip\z@{##}$&\global\@eqcnt\@ne
\hfil$\displaystyle{{}##{}}$\hfil
&\global\@eqcnt\tw@$\displaystyle\tabskip\z@{##}$\hfil
\tabskip\@centering&\llap{##}\tabskip\z@\cr}
\makeatother

\begin{document}
\setlength{\baselineskip}{15pt}
\title{\bf{The flux phase problem on the ring}
}
\author{
Fumihiko Nakano
\thanks{
present address:
Department of Physics, Princeton University, 
Princeton, NJ08544, USA, 
On leave of absence from:
Mathematical Institute, Tohoku University, 
Sendai, 980-77, Japan.} 
}
\date{}
\maketitle

\begin{abstract}
We give 
a simple proof to derive the optimal flux 
which minimizes the ground state energy 
in one dimensional Hubbard model, 
provided the number of particles is even. 

\hspace*{1ex}


\end{abstract}

\hspace*{1ex}

\section{Introduction}
\hspace*{3ex}
We consider 
the Hubbard model on the ring 
({\it i.e.}, 
one dimensional system with periodic 
boundary condition), 
where the magnetic flux is threaded 
through the ring. 
Our problem is 
to obtain the optimal flux 
which minimizes the ground state energy. 
To be precise, 
we define 
the Hubbard Hamiltonian as follows:
%
%
\begin{eqnarray*}
H:=
\sum_{\sigma=\uparrow, \downarrow}
\sum_{x=1}^L
t_{x,x+1}c_{x+1,\sigma}^{\dagger}
c_{x,\sigma}
+
(h.c.)
+
\sum_{x=1}^L
U_x n_{x,\uparrow}n_{x,\downarrow},
\end{eqnarray*}
where 
$L$ ($L\geq 3$) is the number of sites, 
the site 
$L+1$ is equivalent to 
the site $1$, 
$t_{x,x+1}\in {\bf C}$, 
$|t_{x,x+1}| \ne 0$, 
$U_x \in {\bf R}$, 
$c_{x,\sigma}^{\dagger}$ 
(resp. $c_{x,\sigma}$) 
is the creation 
(resp. annihilation) 
operator 
which satisfies 
the canonical anticommutation relations, 
and 
$n_{x,\sigma}:=
c_{x,\sigma}^{\dagger}
c_{x,\sigma}$. 
We write 
$t_{x,x+1}=|t_{x,x+1}|\exp[i\theta_{x,x+1}]$, 
$\theta_{x,x+1} \in [0, 2\pi)$. 
Then, the flux 
which penetrates the ring is defined to be 
$\varphi:=\sum_{x=1}^L \theta_{x,x+1}$. 
The ground state energy 
$E$ 
(in some fixed number of particles $N_e$) 
can be regarded as a function 
of $\varphi$
(and hence we write $E=E(\varphi)$), 
because it does not 
depend on any choice of 
$\{\theta_{x,x+1} \}_{x=1}^L$ 
which satisfies 
$\sum_{x=1}^L \theta_{x,x+1}=\varphi$. 
Our aim is 
to obtain the flux $\varphi = \varphi_{opt}$ 
which attains 
$\min_{\varphi\in[0,2\pi)} E(\varphi)$. 
We call $\varphi_{opt}$ 
as the optimal flux. 
In general, $\varphi_{opt}$ 
is not unique, 
and we will not discuss the uniqueness 
question in this paper. 
There are 
some closely related problems 
in the literature 
(our problem is the same as 
mentioned in (3) below). 
(1) 
it appears 
in a theory of 
superconductivity [AM, W], 
(2) 
In the 
study of the persistent current 
[FMSWH, K, YF, FK], 
they discussed whether the response 
of the Hubbard ring to the external 
field is diamagnetic or paramagnetic, 
and the influence of the electron-electron 
interaction to this property, 
(3) 
In high 
dimensional lattice, the flux phase conjecture 
[HLRW] says that the optimal flux 
per plaquette is equal to the particle  
density per site. 
This implies that 
the diamagnetic feature, 
which widely holds in the one particle  
system, 
becomes opposite in high electron density 
regime. 
This conjecture 
was rigorously proved by Lieb [L] 
at half filling. 
Macris-Nachtergaele [MN] 
gave an improved proof of [L]. 

As for 
the rigorous study of the Hubbard ring 
(of even length), Lieb-Loss [LL] considered 
free electron case ($U_x \equiv 0$) 
at half filling, 
and computed $\varphi_{opt}$ in  
general situation so that translation 
invariance is not assumed. 
They also 
considered what have more complicated 
geometry such as tree of ring, ladder, 
etc.
Lieb-Nachtergaele [LN] 
computed $\varphi_{opt}$ also at half 
filling when $U_x \equiv U$ is any 
constant. 
In this paper, 
we obtain $\varphi_{opt}$ 
when $U_x$ and $L$ are arbitrary, 
while $N_e$ is even. 
Due to the 
hole-particle symmetry, 
it suffices to consider 
$N_e \leq L$. 
 
\vspace*{1em}

{\bf Theorem}

{\it
Let $N_e (\leq	L)$ be even. 

(1) 
Assume 
$U_x < +\infty$ for all $x$. 
$E(\varphi)$ is 
minimized if $\varphi \equiv (N_e/2 + 1)\pi$ 
(mod $2\pi$) 
(resp. $\varphi \equiv N_e \pi/2 $) 
when $L$ is even (resp. $L$ is odd).

(2)
When $U_x = \infty$ for all $x$, 
$E(\varphi)$ is minimized 
if $\varphi=0$, $\pi$. 

}

\vspace*{1em}

%

{\bf {\it Remarks.}}

(1) 
We can derive 
the optimal flux in $S^z \ne 0$ subspaces. 

(a) 
$U_x< +\infty$:
the optimal flux takes $0$ and $\pi$ 
alternatively as $S^z$ varies. 
For instance, 
when $N_e=4n$, and $L$ is even, then 
$\varphi_{opt}=\pi$ 
($S^z=0, 2, 4, \cdots)$, 
and 
$\varphi_{opt}=0$ 
($S^z=1, 3, 5, \cdots)$. 

(b)
$U_x \equiv \infty$:
let $m:=N_{\uparrow}/N_{\downarrow}$ 
($N_{\uparrow}$ (resp. $N_{\downarrow}$) 
is the number of up (resp. down) spins). 
We suppose 
$N_{\uparrow} \ge N_{\downarrow}$ here. 
When $m \notin {\bf N}$, 
$\varphi_{opt}=2k\pi/N_e$, $k\in {\bf Z}$ 
(in this case, 
particles can also be regarded as 
hard core bosons). 
When $m \in {\bf N}$, 
$\varphi_{opt}=
2k\pi/(m+1) - (N_e-1)\pi$ (if $(m+1)L$ is even), 
and 
$\varphi_{opt}=
(2k-1)\pi/(m+1) - (N_e-1)\pi$ (if $(m+1)L$ is odd), 
$k\in {\bf Z}$. 

(2)
When $U_x \equiv +\infty$, 
the proof of Theorem tells us that 
the ground state energy is periodic 
w.r.t. $\varphi$ 
with period $\pi$ (when $S^z=0$), 
and period $2\pi/N_e$ (when $m \notin {\bf Z}$). 
This fact 
and its implications 
are discussed by Kusmartsev and 
Yu-Fowler[K, YF]. 

(3)
When $N_e$ is odd 
and $U_x \equiv +\infty$, 
we can still derive the optimal flux, 
and the result is the same as stated in 
Remark (1). 

On the other hand, 
when $U_x <+\infty$, and $N_e=L$ (half-filling), 
we believe $\varphi_{opt}=\pi/2$, $3\pi/2$ 
as some examples imply 
({\it e.g.}, 
take $t_{x,x+1}$: constant and $U_x \equiv 0$). 
However, in general cases, $\varphi_{opt}$ 
could be different 
depending on the value of $U_x$.
For example, 
let $L=4$, $N_e=3$, and $t_{x,x+1} \equiv t$. 
When $U_x \equiv 0$, 
$E(\varphi)$ is minimized 
if and only if $\varphi=\pm 4 \arcsin (1/\sqrt{5})$, 
while in case of $U_x \equiv +\infty$, 
$E(\varphi)$ is minimized 
if and only if 
$\varphi=0$, $2\pi/3$, $4\pi/3$. 

(4)
$SU(2)$ invariance 
as well as translation invariance
is not necessary to prove Theorem. 
We can let $t_{x,x+1}=t_{x,x+1}^{\sigma}$ 
($\sigma=\uparrow$, $\downarrow$) 
depend also on spin variable. 
In this case, our theorem mentions the optimal flux 
in the $S^z=0$ subspace only. 
Besides, 
our Hamiltonian can include the one body 
potential term as well. 

(5)
The argument 
in the proof, together with that in [LM] 
yields the ground state is unique 
and has spin zero: $S=0$, 
provided the flux $\varphi$ takes the value 
as stated in Theorem.  
Moreover, 
if we let $E(S)$ 
denote the ground state energy in spin $S$ 
subspace, then we have 
$E(S) < E(S+2)$. 
It becomes 
equality when $U_x \equiv \infty$, 
$\varphi=\pi$ (resp. $\varphi=0$), 
and 
$L$: even (resp. $L$: odd).  

(6)
In general, 
the spin of the ground state is sensitive 
to the flux. 
For example, 
let $L$ be even, $N_e = 4n+2$, $t_{x,x+1}\equiv t$, 
and $U_x \equiv \infty$. 
Then, one can show that, 
(a) 
when $\varphi=0$, 
there is a singlet ground state, 
but no ferromagnetic ones. 
(b)
when $\varphi=\pi$, 
there is a ferromagnetic ground state. 

(7)
When $U_x = \infty$, 
not for all $x$, 
the argument of the proof says the following: 
if $\sharp \{x: U_x =\infty \}
\leq L-N_e/2$, 
then the result is the same as in Theorem (1). 
Otherwise, 
the result is the same as in Theorem (2). 

(8)
When $U_x \equiv 0$, 
and $t_{x,x+1} \equiv t$, 
$E(\varphi)$ is 
maximized if and only if 
$\varphi \equiv N_e \pi/2$ 
(resp. $\varphi \equiv (N_e /2 + 1)\pi$), 
if $L$ is even (resp. $L$ is odd), 
which  
should be compared with 
the fact that 
$E(0) = E(\pi)$ 
when $U_x \equiv \infty$ 
(Theorem (2)). 

(9)
When we 
let $L$ large, 
$| E(0) - E(\varphi)|$ 
will behave as $O(1/L)$ [LN]. 

\vspace*{1em}

In section 2, 
we give the proof of Theorem, which is very simple. 
Our problem is 
reduced to consider an one-particle Hamiltonian 
${\cal H}(\varphi)$ on the graph $G$ 
which is composed of the basis of $N_e$-fermion 
Hilbert space. 
Theorem follows 
from obtaining the optimal flux of 
${\cal H}(\varphi)$ on $G$, 
by the usual diamagnetic inequality 
argument. 
%

%
%

\section{Proof of Theorem}
\hspace*{3ex}
At first, 
we consider the case in which  
$L$ is even and $N_e=4n$. 
Due to the 
$SU(2)$ invariance, it is sufficient to 
work on $S^z=0$ subspace 
({\it i.e.}, $N_{\uparrow}=N_{\downarrow}=2n$). 
We fix 
the basis of the Hilbert space 
of $N_e$-fermions: 
\begin{eqnarray*}
{\cal B}:=
\{
\,
c_{x_1,\sigma_1}^{\dagger}
c_{x_2,\sigma_2}^{\dagger}
\cdots
c_{x_{N_e},\sigma_{N_e}}^{\dagger}
&|&{\rm vac}>
\,
:
\\
x_1 \le x_2 \le \cdots \le x_{N_e}, 
&&\,
\sigma_i=\uparrow, \downarrow, 
\,
i=1, \cdots, N_e
\},
\end{eqnarray*}
that is, 
to arrange particles in increasing order 
w.r.t. the space coordinates. 
Our problem is 
equivalent to consider the one-particle 
Hamiltonian ${\cal H}(\varphi)$ on the graph $G$ 
whose sites are composed of ${\cal B}$. 
\[
({\cal H}(\varphi) u)(x):=
\sum_{y\in{\cal B}}s_{xy}(\varphi)u(y), 
\]
where 
$s_{xy}(\varphi):=
<x|H|y>$, 
$x$, $y\in{\cal B}$. 
Two sites 
$x$, $y\in G$ are connected by a bond 
if and only if $s_{xy}(\varphi)\ne0$
(we note 
$|s_{xy}(\varphi)|$ does not depend on 
$\varphi$). 
For given 
$\varphi \in [0, 2\pi)$, 
we fix some $\{ \theta_{x,x+1} \}_{x=1}^L$ 
such that 
$\sum_{x=1}^L \theta_{x,x+1} =\varphi$, 
and thus we suppose 
$s_{xy}(\varphi)$, and hence 
${\cal H}(\varphi)$, is determined by $\varphi$. 
Then, 
it is not hard to show that:
(1) 
every circuit in $G$ has even length 
(because $L$ is even), 
(2)
the fluxes in these circuits 
are always integer multiple of 
$\psi:=
\varphi + 2n\pi + (4n-1)\pi$. 
In fact, 
let ${\cal C}$ be the set of circuits in $G$ 
which have minimal length. 
Every elements 
of ${\cal C}$ is given by fixing 
all particles which have down (resp. up) 
spins and moving each up (resp. down) spins 
all together until each spins come 
to next spin. 
To make it clear, 
we write down an element of ${\cal C}$
when $L=N_e=4$:
\begin{eqnarray*}
c_{1,\uparrow}^{\dagger}
c_{2,\downarrow}^{\dagger}
&c_{3,\uparrow}^{\dagger}&
c_{4,\downarrow}^{\dagger}
|{\rm vac}>
\quad
\leftarrow
\quad
c_{2,\downarrow}^{\dagger}
c_{3,\uparrow}^{\dagger}
c_{4,\uparrow}^{\dagger}
c_{4,\downarrow}^{\dagger}
|{\rm vac}>
\\
&\downarrow&
\qquad\qquad\qquad\qquad\qquad
\uparrow 
\\
c_{2,\uparrow}^{\dagger}
c_{2,\downarrow}^{\dagger}
&c_{3,\uparrow}^{\dagger}&
c_{4,\downarrow}^{\dagger}
|{\rm vac}>
\quad
\rightarrow
\quad
c_{2,\uparrow}^{\dagger}
c_{2,\downarrow}^{\dagger}
c_{4,\uparrow}^{\dagger}
c_{4,\downarrow}^{\dagger}
|{\rm vac}>
\end{eqnarray*}
The second term 
$2n\pi$ in the definition of $\psi$ 
comes from the fact 
that up spins jump down spins 
$2n$ times on the above process, 
and each jump causes to put $(-1)$ 
on the corresponding $s_{xy}(\varphi)$. 
The third term 
$(4n-1)\pi$ in the definition of 
$\psi$ comes from the fact that 
the $2n$-th up spin jumps 
all the other $(4n-1)$ particles 
when it moves from the site $L$ to the site $1$, 
because we set the basis such that 
particles are arranged 
in increasing order.  

On the other hand, 
because of the inequality:
$\sum_{x,y\in{\cal B}}
s_{xy}(\varphi)
\overline{u(x)}u(y)
\ge
-\sum_{x,y\in{\cal B}}
|s_{xy}(\varphi)|
|u(x)||u(y)|$, 
we know that 
the ground state energy is minimized 
when all off-diagonal elements 
$s_{xy}(\varphi)$, $x\ne y$, 
are non-positive. 
Let 
$({\cal H}_{-}u)(x):=
-\sum_{y\in{\cal B}}
|s_{xy}(\varphi)|u(y)$. 
When $\psi\equiv 0$ (mod $2\pi$), 
${\cal H}(\varphi)$ is unitarily equivalent 
to ${\cal H}_{-}$, 
because the fluxes of all circuits in $G$ 
are all the same [LL, Lemma 2.1]. 
$\psi \equiv 0$ (mod $2\pi$) 
yields $\varphi\equiv \pi$ (mod $2\pi$). 
This concludes the proof 
when $L$ is even and $N_e=4n$. 

When $L$ is even 
and $N_e=4n+2$, the only thing we have to do is 
to replace $\psi$ in the above argument by 
$\psi':=
\varphi + (2n+1)\pi + (4n+1)\pi$. 
When $L$ is odd, 
then $\psi$ 
(or $\psi'$ in case of $N_e=4n+2$), 
should satisfy 
$\psi\equiv\pi$ (mod $2\pi$) 
to have optimal flux, 
because 
the minimal length of 
the circuits in $G$ 
is odd, 
and so, the flux of ${\cal H}_{-}$ on 
every elements of ${\cal C}$ 
is $\pi$. 
When $U_x \equiv \infty$, 
the minimal length of circuits in $G$ 
is $2L$, whose 
flux is $2 \varphi + 2 (N_e-1)\pi 
\equiv 2 \varphi$ (mod $2\pi$). 
$\Box$

\vspace*{1em}

{\it Remarks}

(1) 
As an 
alternative proof, one can compute 
the partition function 
$P(\varphi):=Tr[\exp(-\beta H)]$ 
by using the path integral representation 
[AL], 
and show that $P(\varphi)$ is maximized 
if $\varphi$ takes the value stated in Theorem. 
This approach 
has been done by [GMMU], 
where they derived the optimal flux 
in the Falicov-Kimball model. 

(2)
When the number 
of electrons is odd, 
the fluxes of elements of ${\cal C}$ 
are different from each other, 
depending on which spins move 
in the circuit. 
For example, let 
$L=N_e=2n+1$, $N_{\uparrow}=n$, 
and $N_{\downarrow}=n+1$. 
By the 
hole-particle transformation only for down spins, 
we can suppose $N_{\uparrow}=N_{\downarrow}=n$, 
but now the flux of down spins is $\pi-\varphi$  
(this situation 
is similar to that discussed in [FK]). 

Our supposition 
is the following: 
the ``contribution" to the ground state energy 
from ${\cal C}$ would cancel
each other, 
and an important contribution would come from 
those circuits where 
up spins and down spins move together 
in the opposite direction 
which has flux 
$\varphi-(\pi-\varphi)=2\varphi-\pi$, 
and has length $2n$ in $G$ 
(the meaning of ``contribution" 
could be clear if we consider 
$Tr[\exp(-\beta H)]$ 
instead of the ground state energy). 
$2\varphi-\pi\equiv 0$ 
would give the minimizing energy. 
However, 
this supposition would 
not be easy to prove. 

(3)
The proof 
above relies on the special nature 
of the ring geometry: 
there is always fixed number of particles 
on only one loop, so that all circuits 
on the graph $G$ favor the same flux 
$0$ or $\pi$, depending on cases. 
However, 
on more complicated systems such as 
two dimensional lattice, 
the graph $G$ has so many different 
circuits which favor different fluxes 
so that our argument does not work 
even if $U_x \equiv \infty$, 
except the Nagaoka-case 
($N_e=|\Lambda|-1$, $U_x \equiv \infty$ [N,T]), 
where the optimal flux is zero everywhere.  

%
%

\section{Conclusion}
\hspace*{3ex}
In this paper, 
we derived the optimal flux $\varphi_{opt}$ 
in the Hubbard model on the ring. 
Our result 
is true in general situation so that the 
translation invariance is not necessary 
to assume, except the number of particles 
must be even. 
In this section, 
we briefly discuss the physical interpretation 
of our result.

The result 
(1) of our theorem is consistent 
with that of [FMSWH], where it is shown 
that, at half-filling, the current response 
of the ground state 
is paramagnetic (resp. diamagnetic) 
when $N_e = 4n$ (resp. $4n+2$) 
by numerical computation. 
However, 
these are not equivalent, 
especially 
when $N_e=4n$. 
In fact, 
[FMSWH] showed, 
when $L=6$, $N_e=4$, and $U_x >0$, 
the ground state is diamagnetic 
(this also implies 
why it is not easy to seek 
$\varphi$ which maximizes $E(\varphi)$). 
Therefore, 
our contribution may be that 
there would be no effects of spatial 
disorder. 

The result (2) 
of our theorem and Remark (2) 
after that is already found and discussed by 
[K, YF]. 
However, our proof gives a different 
picture: 
the graph $G$ consists of rings 
of larger lengths, 
for $U=\infty$ prohibits the exchange 
of particles. 

Finally, 
our argument gives a ring version of 
the Lieb-Mattis theorem [LM] 
when $\varphi=\varphi_{opt}$ 
(Remark (5) after Theorem). 

%

\vspace*{1em}

{\bf Acknowledgement}

The author 
would like to thank 
professor E. H. Lieb 
for pointing out Remark (4)(5). 
The author is 
partially supported by the 
Japan Society for the Promotion of Science. 

\vspace*{1em}


\end{document}